\documentclass[epj]{svjour}
\usepackage{graphicx}
\usepackage{graphics}
\def\be{\begin{equation}}
\def\ee{\end{equation}}
\def\bee{\begin{eqnarray}}
\def\eee{\end{eqnarray}}

\begin{document}
\title{Emission of cosmic rays from Jupiter. Magnetospheres as possible Sources of Cosmic Rays}
\author{G.Pizzella}% etc
%}                     % Do not remove
%
\offprints{}          % Insert a name or remove this line
\institute{Istituto Nazionale di Fisica Nucleare, Laboratori Nazionali 
di  Frascati}
\date{Received: date / Revised version: date}
\abstract{
Measurements made recently with the PAMELA satellite have shown with good evidence that a fraction of the cosmic rays detected on Earth comes from Jupiter. This result draws attention to the idea  that magnetospheres of astrophysical objects could contribute to the sources of cosmic rays. We discuss this conjecture on the basis
 of Earth installed instrumentation and of the measurements made with PAMELA.
The experiments strongly favor the validity of the conjecture. In particular the PAMELA data show that the proton fluxes are larger when the Earth orbit intersects the lines of the interplanetary magnetic field connecting Jupiter with Earth. This effect shows up with more than ten standard deviations,  difficult to explain without the idea that  part of the  cosmic ray protons comes directly from the Jupiter magnetosphere. 
 % end of PACS codes
} %end of abstract
\maketitle
%

%\begin{document}
\section{Introduction}

It has been recently published a paper \cite{pamela1} showing good evidence that a small but not negligible  fraction of cosmic rays detected by the Earth satellite PAMELA\footnote{PAMELA, launched on 15th June 2006, in a 350-600 km heigh orbit, is a space-based experiment designed for precise measurements of charged cosmic rays - protons, electrons, their antiparticles and light nuclei  in the kinetic energy interval from several tens of MeV up to several hundreds of GeV \cite{pamela}.} comes from the planet Jupiter. The suggestion was that higher energy particles leak out from the Jupiter trapped particle region and then  are ejected into space following the interplanetary magnetic field (IMF) lines of force, eventually reaching Earth. 

This result draws attention to the idea put forward long ago \cite{pizz} that magnetospheres of astrophysical objects  could contribute to the sources of cosmic rays.

On the other hand it is believed today that pulsars  be one of the candidates of the observed ultra-high-energy cosmic rays,  getting power by a centrifugal mechanism of acceleration. The result obtained with the cosmic rays coming from the Jupiter magnetosphere reinforces the possibility that astrophysical magnetosphere be indeed the source of the cosmic radiation.

In this paper we wish to exploit the experimental result and investigate more thoroughly  the above suggestion.

\section{The trapped radiation}

An unexpected discovery was the Van Allen belt in 1958. Although  in earlier times Kristian Birkeland, Carl St\"{o}rmer, and Nicholas Christofilos considered the possibility that fluxes of charged particles could dwell in the magnetic field surrounding the Earth, nobody was expecting so large density of charged particle population that formed a sort of permanent belt  surrounding the Earth. As matter of fact when Explorer 1, under the leadership of James A. Van Allen, reached high altitudes, the Geiger counters designed for much lower flux of particles stopped to operate and it took the wisdom of Carl E. McIlwain, who had calibrated the detectors, to understand that the fluxes were so large to saturate the apparatus. 

The natural question is how the Van Allen belt is originated. Since the discovery, it has been suggested that the high energy trapped protons be generated by solar neutrons  decaying just in the Van Allen region, but this is not sufficient to explain the high variability of the trapped radiation correlated with the flurries of the solar wind (see ref. \cite{forbush}). Very many papers have been written during the sixty year from the discovery of the Van Allen belt, investigating the behaviour of the trapped particles.

Electromagnetic waves have been detected since the beginning (see \cite{waves}), presumably originated by the action of the solar wind on the magnetosphere \cite{waves1} and this justifies the presence of acceleration mechanisms which act on the magnetospheric particles.

One important point is to understand whether the particle-wave interaction accelerate particle to the very high energy measured in the radiation belt.

This problem  has been debated since their discovery in 1958. More recently  Reeves et al. \cite{reeves}  used data from the Van Allen Radiation Belt Storm Probes, launched by NASA on 30 August 2012, discovered that radiation belt electrons are accelerated locally by wave-particle interactions

The data showed an increase in energy that started right in the middle of the radiation belts and gradually spread both inward and outward, implying a local acceleration source. This local energy comes from electromagnetic waves coursing through the belts, tapping energy from other particles residing in the same region of space.

  The authors quote: \it the measurements show signatures of local acceleration by wave particle interactions in the heart of the radiation belts.\rm

Recent results by the Van Allen Probes mission showed that the occurrence of energetic ion injections inside geosynchronous orbit could be very frequent throughout the main phase of a geomagnetic storm \cite{storm}. Thus it is the solar wind which powers the local acceleration mechanism,

Many experiments by means of several spacecraft (Pioneer 10 and 11, Voyager 1 and 2, Ulysses. Galileo, Cassini) have shown that all major planets in the solar system possessing a magnetic field also have Van Allen belts surrounding them \cite{JAVA}, in particular Jupiter which has a dipole magnetic field about two thousand times greater than that of the Earth and a trapping region containing very high energy particles.

In the Jovian radiation belts, trapped particles are about ten times more energetic than the ones from the equivalent radiation belts of Earth. In addition, they are several orders of magnitude more abundant\cite{progiove}. Indication of accelerating mechanisms are also given \cite{acce,teegarden}.

In this paper we want to study how magnetospheric systems, home of accelerating mechanisms, contribute to the cosmic radiation that pervades the Universe. 

\section{Cosmic rays}

Gold \cite{gold} suggested that cosmic rays be generated by the explosion of supernovae, but it is likely that there can be a variety of sources like, i.e. active galactic nuclei and, as suggested by Fermi \cite{fermi}, sources of continuos nature acting in the intergalactic space. 

The observation \cite{auger} of ultrahigh-energy cosmic rays (UHECR), with
energies above $6\times 10^{19}$ eV, obtained with the Pierre Auger Observatory during 3.7
years, has shown that these cosmic rays have a non-isotropic spacial distribution.
Quoting paper: \it the highest-energy particles originate from nearby extragalactic
sources whose flux has not been substantially reduced by interaction with the
cosmic background radiation. The Collaboration suggests that AGN or similar objects could be possible sources.\rm

It is reasonable to think that if the Fermi theory applies also to the UHECR,
one should expect an isotropic distribution for them, since the acceleration
mechanism should operate everywhere in the intergalactic space. This is not verified by the Auger experiment.

In the following we discuss a new conjecture about the cosmic ray sources:
 the  \it Magnetospheric Cosmic Ray Conjecture  \rm (MCRC). 
 
 The idea is:  particles trapped in the magnetic field of a magnetosphere, under the action of some acceleration mechanism, leak out from the trapping region when their motion violate the adiabatic invariance. This occurs when  the Alfven condition (see also the Hillas criterion \cite{hillas})
 \be
 \frac{dB/dr)R_L}{B}\leq\epsilon
 \ee	
  is not satisfied anymore	\cite{alfen,break}					
 (B is the magnetic field, $R_L$ the Larmor radius and  $\epsilon$ is a dimensionless quantity).
 
A prediction of this model  is a cosmic  cutoff  energy  given by
\be
E_{max}=\frac{\epsilon cqM}{3r_S^2}
\label{maxx}
\ee
where $r_S$ is the radius of the source and M its magnetic dipole. 

In the Van Allen belt of Earth with $M\equiv M_E= 8\cdot 10^{15}~T m^3$, the largest energy of protons is of the order of a few hundred MeV. Pamela \cite{pamela} measured about 1 GeV  in the shadow zone.

Although the largest energy might depend on other parameters, in addition to those appearing in Eq.\ref{maxx}, for Jupiter, with $r_S\sim 71500~ km$ and $M_J \sim 1800\times M_E $,  we try to scale up from the Earth 500 MeV, obtaining 
\be
E_{max}\sim 70~GeV
\label{giove}
\ee
This is a very upper limit for Jupiter, greater than the observed proton energy in the Jovian magnetosphere. Because of the complexity of the Jovian magnetosphere (including the effect of its Io satellite) with respect to the Earth's one, it is likely that the real upper limit be smaller.

In the case of cosmic rays of extra solar-system origin, for a neutron star with $r_S\sim 10~ km$ and $M_{NS}= 10^{21}~T m^3$, we calculate
\be
E_{max}\sim 3\cdot 10^{19}~eV
\ee
This value compares with the highest cosmic ray proton energy detected so far. Higher values can be obtained for larger values of M.

\section{Experimental evidence validating MCRC}

The planet Jupiter, possessing an enormous magnetosphere, is well apt to test MCRM.  

If charged particles are accelerated in the magnetosphere and then escape according to MCRC, they should follow the interplanetary magnetic field  lines\footnote{This is possible, because 1 GeV protons have a Larmor radius,  for a IMF between Jupiter and Earth of about 1 nT,  of the order of  2\% of AU.} finally reaching Earth where they are detected within the cosmic rays fluxes.

In the following we start by considering the Interplanetary Magnetic Field (IMF) as measured by interplanetary spacecrafts and then we study the cosmic rays measured on Earth and aboard the PAMELA satellite.

\subsection{The Interplanetary Magnetic Field}

The IMF  has been investigated with various spacecrafts (Helios 2, IMP8,
Pioneer, Venus Orbiter, Voyager 1, Ulysses) and
 plays an important role, because its  lines of force drive the protons with energy  up to a few GeV  from their possible source, Jupiter, to Earth. The IMF originates in the Sun and is transported by the solar wind, that travels with velocity $V_R$. If the Sun did not rotate, the magnetic lines would have been radial straight lined. Due to the Sun rotation ($\Omega\sim$ 25 days) the lines become spirals around the Sun \cite{fitz}.

The equation of the magnetic field lines in the ecliptic  plane
can be derived from  the equation
\be
\frac{rd\phi}{dr}\simeq \frac{B_t}{B_r}\frac{V_R}{\Omega}=0.9 \frac{V_R}{400~ km/s} \frac{B_t}{B_r}
\ee
where the distances are expressed in AU and the angles in radians.

The magnetic lines of force obtained by integrating this equation have the shape of spirals. This is verified by the experiments with space borne magnetic field detectors,  all giving an IMF with spiral lines of force, but the precise behavior depends on the radial gradients of the ratio $B_t /B_r$, that, in the Parker theory, scales linearly with $r$.
Measurements made, in particular, with the Pioneer 11, Voyager and Ulysses spacecrafts \cite{olga,ulisse,kenneth,smith3,winter,smith1,kharabova} show some discrepancies from the behavior predicted in the Parker theory. 

Let us consider the angle  $\Phi_{EJ}=\overbrace{Jupiter. Sun .Earth}$, the angle 
between Jupiter, the Sun and the point where the IMF line intersect the Earth orbit.

Putting, as already suggested in \cite{kenneth}, 
$
B_t/B_r\sim r^\eta
$
we find the  magnetic line of force connecting Jupiter with Earth
\be
\Phi_{EJ}=\phi_J-\phi_E=\frac{0.9}{\eta}(5.2^\eta-1^\eta)\frac{V_R}{400~ km/s}
\ee
where $\eta=1$ in the Parker theory.

We show in the Table \ref{fi} the angle $\Phi_{EJ}$ for the Parker model and that obtained from the measurements made with the space probes
\begin{table}[hbtp]
\begin{center}
\caption{The angle $\Phi_{EJ}$ for $V_R=400~ km/s$ .}
\vspace{3 mm}
\begin{tabular}{|ccc|}
\hline
$\eta$&$\Phi_{EJ}$&reference\\
\hline
1&$216^o$&Parker model\\
0.57&$140^o$&Khabarova\cite{olga,kharabova}\\
0.37&$117^o$&Behannon \cite{kenneth}\\
\hline
\end{tabular}
\label{fi}
\end{center}
\end{table}

In the past, using the experimental IMF as measured by the above spacecrafts,   it has been calculated \cite{add} $\Phi_{EJ}\sim95^o - 134^o$.

\begin{figure}[hbtp]
\includegraphics[width=1.0\linewidth]{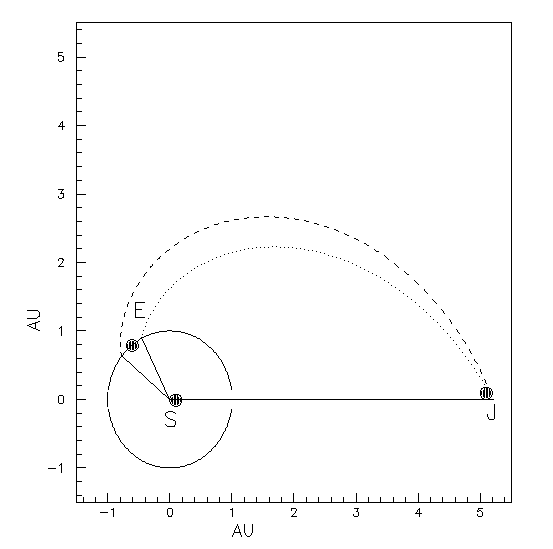}
 \caption{
A schematic view of Sun, Earth and Jupiter  with the IMF lines connecting Jupiter with  Earth. The dashed line refers to $\eta=0.57$, the point line to $\eta=0.37$. We notice that the IMF lines intersect the Earth orbit at angles $\Phi_{EJ}<180^o$.
       \label{parker} }
\end{figure}

A schematic view of Sun, Earth and Jupiter  with the IMF lines connecting Jupiter with  Earth is shown in Fig.\ref{parker}.
 
\subsection{Experiments with Earth based instrumentation}

Earth based instrumentation  has been considered for studying whether was possible that an even  small fraction of the cosmic rays detected by this instrumentation came from Jupiter.

The first attempt was made with the available neutron monitors. It was found \cite{venditti} that the measured cosmic ray flux
was enhanced at certain times when considered in a reference system coo-rotating with Jupiter. The enhancement was of the order of 0.4\%
and occurred for values  $\Phi_{EJ}\sim 60^o-180^o$ with maximum at $\Phi_{EJ}\sim125^o$ .

A similar study was made by B. Mitra et al. \cite{mitra1,mitra}. They found   a strong modulation at the synodic Jupiter period above 99\% confidence level and an angle $\Phi_{EJ}=140^o-160^o$, and concluded  that  there was \it  a strong possibility that a part of cosmic rays  comes from Jupiter\rm.

A negative result was instead found in \cite{swinson} and \cite{naga}. They were focused in particular on the idea that the magnetic field of Jupiter could shield the cosmic rays coming from the outer space. Their conclusion was that \it the Jovian effect on cosmic rays observable at the Earth's ground is minimal\rm.

\subsection{Measurement in the outer space with the PAMELA satellite}
Measurements made in the outer space with instrumentation dedicated to the study of cosmic rays are much more apt to study the problem of a possible Jovian emission. 

This has been done with the PAMELA space-borne instruments  for protons with energy in the range 0.4-50 GeV \cite{pamela}.  About 1800 daily averages  of proton flux were considered and analyzed\cite{pamela1}.

The proton fluxes were measured before and after a solar maximum. In order to get rid of the well-known modulation during the solar cycle the data have been fitted and  the quantity $\xi(t)=protonflux(t)-fit(t)$ was considered in the analysis, as described in \cite{pamela1}.

% We give here a summary and discuss in more detail the statistical meaning of the result.

The reasoning was the following: looking at fig.\ref{parker} we realize that if MCRC is correct, when the  Earth, during its revolution around the Sun, intersects the lines of the IMF which come from Jupiter   at angles $\Phi_{EJ}<180^o$ (for solar wind velocity of less than $550~km/s$) larger fluxes of protons should be expected. Thus, the integrated proton flux with $\Phi_{EJ}<180^o$ ought to  be larger than the integrated flux with $\Phi_{EJ}>180^o$. 

This is indeed what has been found \cite{pamela}, as shown in Table \ref{risultati}.
\begin{table}[hbtp]
\begin{center}
\caption{The second column indicates the difference between the average flux for $\Phi_{EJ}<180^o$ and that for $\Phi_{EJ}>180^o$ (a null value was expected). The distributions of the daily fluxes for the two categories have been checked to be gaussian and the indicated errors are standard deviations.}
\vspace{3 mm}
\begin{tabular}{|ccc|}
\hline
Rigidity&Excess flux D&SNR\\
Interval (GV)&Proton/(cm2 sr s GV)&\\
\hline
0.4-0.65&$(6.09\pm 0.52)\cdot 10^{-4}$&11.7\\
0.65-15&$(5.95\pm 0.60)\cdot 10^{-4}$&9.9\\
15-50&$(1.72\pm 0.41)\cdot 10^{-6}$&4.2\\
\hline
\end{tabular}
\label{risultati}
\end{center}
\end{table}

The  average proton flux for $\Phi_{EJ}$ less than $180^o$  exceeded the average flux for more than $180^o$ by several standard deviations, namely about 12 standard deviations for rigidities in the range 0.4-0.65 GV, about 10 for rigidities in the range 0.65-15 GV and 4 for rigidities in the range 15-50 GV, as shown in the Table \ref{risultati}.

The smaller effect at the higher rigidities could be explained with MCRC that imposes upper limits to the production of cosmic rays by the Jupiter magnetosphere.

The result shows up with so many standard deviations with respect to the null hypothesis to lead us to state that is established  beyond any reasonable doubt.
In ref.\cite{pamela} it is shown that this effect occurs both before and after the solar maximum.

This statement leads us to investigate the correctness of our error estimates.

\begin{figure}[hbtp]
\includegraphics[width=1.0\linewidth]{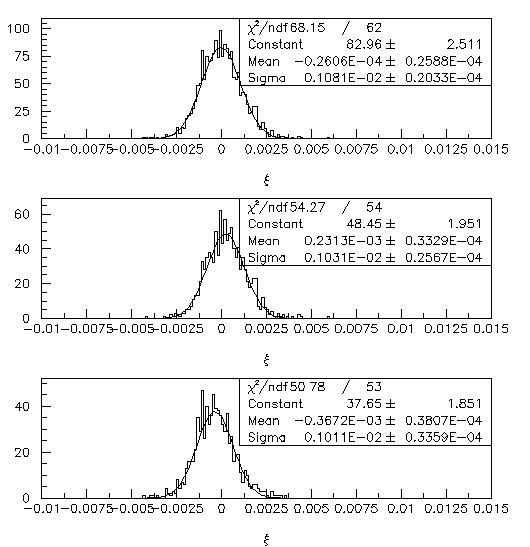}
 \caption{
Rigidity 0.4-0.65 GV.  On the top the gaussian distribution for all 1965 data, in the middle the distribution for 1053 data having $\Phi_{EJ} <180^o$, on the bottom the distribution for $\Phi_{EJ}>180^o$. No daily averages for the entire period were eliminated.
       \label{gauss} }
\end{figure}

In order to give proper meaning to the errors, we show in fig.\ref{gauss} the $\xi(t)$ daily average distributions and their  gaussian fits for the rigidity  0.4-0.65 GV.

We note  that the distribution of the daily averages during the entire period (upper graph) is gaussian and that the average is $(-0.26\pm 0.26)10^{-4}$ (proton/(cm2 sr s GV), that is null, as expected.

We also find that the distributions for the two categories,  $\Phi_{EJ} <180^o$ and  $\Phi_{EJ} >180^o$, are both gaussian. Using the averages of the gaussian fits, we get $D=(0.598\pm0.050)10^{-3}$, that is $D\ne0$ with $SNR=\frac{0.598}{0.050}=12$ standard deviations. Using the averages of the measured daily data we obtain SNR=11.7 (see Table \ref{risultati}).

\section{Discussion and conclusion}

The discovery of the Van Allen belt  in a magnetosphere due to the dipole field of the Earth showed, already in 1958, that  electrically charged particle are concentrated in certain regions of space where acceleration mechanisms operate. Thus the MCRC comes out in a natural way, that the magnetospheres leak out the most energetic particles. 

The vicinity of a powerful magnetosphere, as that possessed by the planet Jupiter, has allowed initially to test MCRC  by studying the data of the Earth based cosmic ray detectors. This test gave indication  that part of the cosmic rays observed on Earth appeared to come from Jupiter. But the real test  has been made with PAMELA,  cosmic ray detector in space.  The  MCRM has been confirmed with over ten standard deviations.

The Jovian result reinforces the theory that pulsars  be one of the candidates of the observed ultra-high-energy cosmic rays and makes
reasonable to extrapolate  this feature to all astrophysical magnetospheres where powerful acceleration mechanisms operate (say:  Pulsars, AGN...). 

\section{Acknowledgments}
I thank Piergiorgio Picozza and his reserch group for the  PAMELA data. I am indebted with Guido Barbiellini and Olga Khabarova for suggestions and discussions and with Carl E. McIlwain and Marcello Piccolo for suggestions.

\end{document}